\documentclass[10pt,a4paper,twocolumn]{article} 

\usepackage{graphicx}
\usepackage{dcolumn}
\usepackage{bm}
\usepackage{gensymb}
\usepackage{amsmath}
\usepackage[affil-it]{authblk}
\usepackage{abstract}
\usepackage[english]{babel} 
\usepackage[hmarginratio=1:1,top=20mm,left=20mm,columnsep=20pt]{geometry} 
\usepackage[noadjust]{cite}

\title{\vspace{-1.1cm}\Large\textbf{Critical behavior in porous media flow}}

\author[1]{M. Moura\thanks{Corresponding author (marcel.moura@fys.uio.no)}}
\author[1]{K. J. M\aa{}l\o{}y}
\author[2]{R. Toussaint}
\affil[1]{Department of Physics, PoreLab, University of Oslo - PO Box 1048, Blindern, N-0316, Oslo, Norway.}
\affil[2]{Institut de Physique du Globe de Strasbourg, UMR 7516, University of Strasbourg - 5 rue Ren\'e Descartes, 67084, Strasbourg, France}

\begin{document}
\twocolumn[
  \begin{@twocolumnfalse}
    \maketitle
  \vspace{-0.9cm}
\begin{abstract}
    \normalsize
The intermittent burst dynamics during the slow drainage of a porous medium is studied experimentally. We have shown that this system satisfies a set of conditions known to be true for critical systems, such as intermittent activity with bursts extending over several time and length scales, self-similar macroscopic fractal structure and $1/f^\alpha$ power spectrum. Additionally, we have verified a theoretically predicted scaling for the burst size distribution, previously assessed via numerical simulations. The observation of $1/f^\alpha$ power spectra is new for porous media flows and, for specific boundary conditions, we notice the occurrence of a transition from $1/f$ to $1/f^2$ scaling. An analytically integrable mathematical framework was employed to explain this behavior.
\vspace{0.5cm}
    \end{abstract}
  \end{@twocolumnfalse}
]
\saythanks

\section{Introduction}
The topic of fluid motion inside a porous network has deservedly been subjected to a considerable number of studies over the past decades. Scientists have studied the morphology and dynamics of the flow \cite{maloy1985,lenormand1989,maloy1992,furuberg1996,grunde2004,toussaint2005,or2008,sandnes2011,moebius2012,deanna2013,macminn2015,bultreys2015} and proposed a set of numerical schemes able to reproduce the observed macroscopic patterns \cite{wilkinson1983,furuberg1988,rothman1990,misztal2015,ferrari2015} and relevant pore-scale mechanisms \cite{haines1930,morrow1970,lenormand1983,martys1991,grunde2005,berg2013,holtzman2015,trojer2015,schluter2016}. The topic is also of central importance for the study of groundwater flows and soil contaminants treatment \cite{guymon1994,jellali2001} and has direct applications in the energy sector, for example, in hydrocarbon recovery methods \cite{tweheyo1999}. One particularly interesting aspect of multiphase flow in porous media is its intermittent dynamics \cite{maloy1992,furuberg1996,haines1930}, with long intervals of stagnation followed by short intervals of strong activity. This kind of general behavior \cite{pomeau1980,pomeau1980_2,hirsch1982} appears in many physical, biological and economical systems, such as the stick-slip motion of a block on an inclined plane \cite{gomes1998}, the propagation of a fracture front in a disordered material \cite{maloy2006,grob2009,tallakstad2011}, the number of mutations in models of biological evolution \cite{sneppen1995}, acoustic emissions from fracturing \cite{sethna2001,stojanova2014}, variations in stock markets \cite{liu1999}, and the rate of energy transfer between scales in fully developed turbulence \cite{kolmogorov1962,salazar2010}. Intermittent phenomena arise irrespectively of the (certainly different) specific details of each system. In the particular case of porous media flows, this is caused by the interplay between an external load (for example, an imposed pressure difference across the system) and the internal random resistance due to the broader or narrower pore-throats.

In the present work we show experimental results on the burst dynamics during drainage in artificial porous media and investigate the question of how the pressure fluctuations (due to the burst activity) can encode useful information about the system. The flows studied are slow enough to be in the capillary regime, in which capillary forces are typically much stronger than viscous ones \cite{lenormand1989.2,maloy1992}. We have employed synthetic quasi-2D systems driven by a controlled imposed pressure (CIP) boundary condition. This boundary condition differs from the controlled withdrawal rate (CWR), more commonly used \cite{maloy1992,moebius2012}. The dynamics is characterized both via direct imaging of the flow and by local pressure measurements. We present results related to the statistics of bursts, their morphology and orientation within the medium, and the power spectral density (PSD) associated with the fluctuations in the measured pressure signal. In particular, we show that for systems driven by the CIP boundary condition, the PSD presents a $1/f$ scaling regime. The presence of $1/f^\alpha$ power spectra is a widespread feature occurring in a myriad of contexts \cite{press1978,mandelbrot1982,schroeder1991}, commonly signaling the collective dynamics of critical systems. Some examples are the early measurements of flicker noise in vacuum tubes \cite{johnson1925}, fluctuations in neuronal activity in the brain \cite{novikov1997}, quantum dots fluorescence \cite{pelton2007}, loudness in music and speech \cite{voss1975,gardner1978} and fluctuations in the interplanetary magnetic field \cite{matthaeus1986}. Although $1/f^\alpha$ power spectra have also been observed in some fluid systems, such as simulations and experiments on hydrodynamic and magnetohydrodynamic turbulence \cite{bourgoin2002,dmitruk2007} and quasi-2D turbulence in electromagnetically forced flows \cite{herault2015}, to the best of our knowledge the results reported here provide the first experimental observations of $1/f^\alpha$ power spectra in porous media flows.

\section{Methodology}

\begin{figure}
	\centering
	\includegraphics[width=\linewidth]{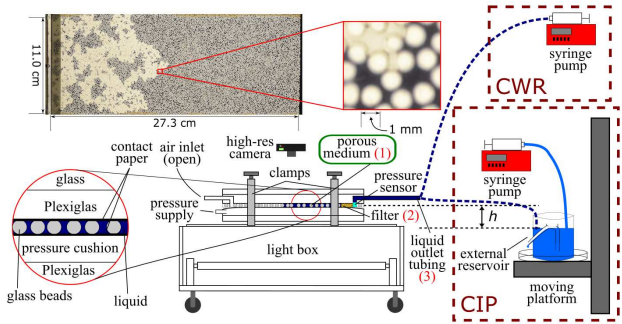}
	\caption{(color online) Diagram of the experimental setup and boundary conditions (CIP or CWR). The numbers (1), (2) and (3) denote the porous medium, filter and external tubing.}
	\label{fig:diagram}
\end{figure}

Fig.~\ref{fig:diagram} shows a schematic representation of the setup employed (additional details in Ref.~\cite{moura2015}). The quasi-2D porous network is formed by a modified Hele-Shaw cell filled with a monolayer of glass beads having diameters $a$ in the range \mbox{$1.0mm<a<1.2mm$}. The beads are kept in place by a pressurized cushion placed on the bottom plate of the cell. A spongeous filter with pores much smaller than those in the medium is placed between the porous network and the outlet of the model. This filter allows the dynamics to continue inside the medium even after breakthrough \cite{moura2015}. Pressure measurements are taken at the outlet with an electronic pressure sensor (Honeywell 26PCAFG6G) that records the difference between the air pressure (non-wetting phase) and the liquid pressure (wetting phase) at the outlet, \textit{i.e.}, $p_m=p_{nw}-p_w^{out}$. Since the inlet is open to the atmosphere, $p_{nw}=p_0$ in all experiments, where $p_0$ is the atmospheric pressure. The porous matrix was initially filled with a mixture of glycerol ($80\%$ in weight) and water ($20\%$ in weight) having kinematic viscosity $\nu = 4.25\;10^{-5} m^2/s$, density $\rho = 1.205\;g/cm^3$ and surface tension $\gamma = 0.064 \;N.m^{-1}$. We have performed experiments on $4$ different porous media with dimensions: (1)~$27.3cm$~x~$11.0cm$, (2)~$14.0cm$~x~$11.5cm$, (3)~$32.8cm$~x~$14.6cm$ and (4)~$32.0cm$~x~$4.5cm$, where the first number corresponds to the length (inlet--outlet direction) and the second to the width. The outlet of the model is connected to an external reservoir. The height difference $h$ between the surface of the liquid in this reservoir and the model is used to control the imposed pressure via an adaptive feedback mechanism (CIP boundary condition). This mechanism guarantees that the pressure is only increased when the system is in a quasi-equilibrium situation (see details in \cite{moura2015}). By slowly increasing the imposed pressure (via small steps in the height of the reservoir $dh=10\mu m \implies dp = \rho g dh = 0.12\;Pa$, where $g$ is the acceleration of gravity), new pore-throats may become available to invasion. The value of $dh$ was chosen to satisfy the accuracy condition that the height would typically have to be increased several times before new pores are invaded. As long as this condition is satisfied, the results obtained should be independent of the particular value of $dh$.

\begin{figure}
	\centering
	\includegraphics[width=\linewidth]{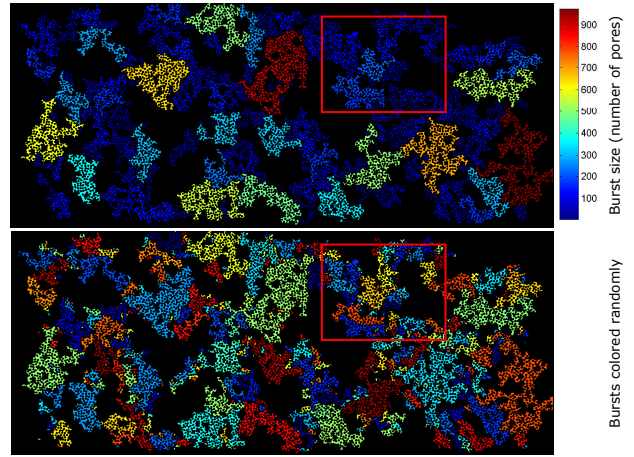}
	\caption{(color online) Individual bursts for experiment CIP-1. The flow is from left to right, during $\approx 82h$. Bursts color coded by their size normalized by a typical pore area (top) and randomly (bottom). The vast blue areas in the top image contain many smaller bursts (detail).}
	\label{fig:burstsarea_random}
\end{figure}

\section{Burst size distribution}

\begin{figure}
	\centering
	\includegraphics[width=\linewidth]{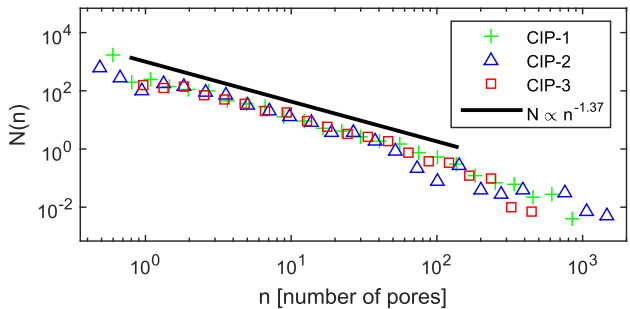}
	\caption{(color online) Burst size distribution $N(n)$. The line shows the scaling $N(n)\propto n^{-\tau}$, with $\tau = 1.37 \pm 0.08$, which is consistent with the theoretical value $\tau=1.30\pm 0.05$ predicted by numerical simulations and percolation theory \cite{martys1991,stauffer1994}. The data has been shifted vertically to aid visualization.}
	\label{fig:burstslocaldist}
\end{figure}

We begin by analyzing the size distribution of invasion bursts in a CIP experiment. A burst is understood as any connected set of pores invaded in the interval $\Theta=t_2-t_1$ between two consecutive time instants, $t_1$ and $t_2$, at which the imposed pressure was increased (i.e., the imposed pressure is constant during the interval $\Theta$, being changed only at its extremes $t_1$ and $t_2$). Fig.~\ref{fig:burstsarea_random} shows the individual bursts for experiment CIP-1 (the number identifies the model), colored according to their area (top) and randomly (bottom), the latter being done to aid the visualization of separate bursts. Only bursts having their centroids in the central $90\%$ of the length are considered, to avoid possible boundary effects \cite{moura2015}. A great deal of information can be obtained from this image. Initially, one can observe the homogeneity and isotropicality of the dynamics: the bursts don't seem  to follow a well defined size gradient (the top image does not seem to transition from blue to red following a specific direction), nor have they a clear preferred orientation (they are not particularly elongated in any direction). It is hard, if not impossible, to say from this image in which direction the invasion takes place (it is from left to right). A reflection (vertical or horizontal) or a $180 \degree$ rotation would also not be clearly identified. The box counting fractal dimension \cite{feder1988,stauffer1994} of the invading cluster was measured to be $D=1.76 \pm 0.05$. Fig.~\ref{fig:burstslocaldist} shows the burst size distribution $N(n)$ for 3 separate experiments (the number of pores $n$ being measured by normalizing the burst area by a typical pore area $\approx 0.3 mm^2$). The system exhibits the scaling $N(n)\propto n^{-\tau}$, with $\tau = 1.37 \pm 0.08$, over at least two decades. The burst dynamics is therefore spatially self-similar, a feature commonly associated with systems close to a critical transition \cite{feder1988,schroeder1991}. The exponent $\tau$ has been calculated via maximum likelihood estimation (MLE) \cite{clauset2009} using the data from Fig.~\ref{fig:burstsarea_random} for burst sizes in the interval $1$~pore~$<n<150$~pores. MLE was used in order to avoid possible biases from data binning (MLE is a binning-free method), see also \cite{iglauer2016}. The scaling is shown in Fig.~\ref{fig:burstslocaldist} on top of the logarithmically binned histogram of the data for the sake of visualization. Experiment CIP-4 was left out of the analysis because boundary effects rendered the results unreliable (model 4 is too narrow). The measured exponent is consistent with the value $\tau=1.30\pm 0.05$ predicted by numerical simulations and percolation theory \cite{martys1991,stauffer1994}. Martys et al. \cite{martys1991} derived the analytical form
\begin{equation}
\tau=1+\frac{D_e-1/\nu'}{D} \:,
\label{eq:tau}
\end{equation}
where $D$ and $D_e$ are respectively the fractal dimensions of the growing cluster and its external perimeter and $\nu'=4/3$ is the exponent characterizing the divergence of the correlation length \cite{feder1988,stauffer1994}. Using the values  $D=1.76$ and $D_e=4/3$ \cite{furuberg1988}, we obtain $\tau=1.33$, very close to the measured value $\tau = 1.37 \pm 0.08$ shown in Fig.~\ref{fig:burstslocaldist}. Our measurements provide a direct experimental verification of Eq.~(\ref{eq:tau}), proposed in Ref.~\cite{martys1991}.

Crandall et al. \cite{crandall2009} performed measurements in a CWR system finding the exponent $\tau = 1.53$, which is compared to the theoretical prediction of $\tau = 1.527$ from Roux and Guyon \cite{roux1989}. Nevertheless, Maslov \cite{maslov1995} pointed out an inconsistency in this theoretical prediction, the correct expression being given in Eq.~(\ref{eq:tau}). Modified invasion percolation simulations and pressure measurements\cite{maloy1992,furuberg1996} have shown that, in a CWR, system very large bursts are split into smaller ones. A burst size distribution was observed, with exponent $\tau = 1.3 \pm 0.05$ for the simulations and $\tau = 1.45 \pm 0.10$ for the experiments (consistent with Eq.~(\ref{eq:tau})), followed by an exponential cutoff \cite{maloy1992,furuberg1996}. In the CIP case large bursts can occur because the displaced liquid can freely flow out of the model but in the CWR case this is not possible since the available volume for the displaced liquid is bounded by the outlet syringe volume.

\section{Burst time distribution}
\begin{figure}
	\centering
	\includegraphics[width=\linewidth]{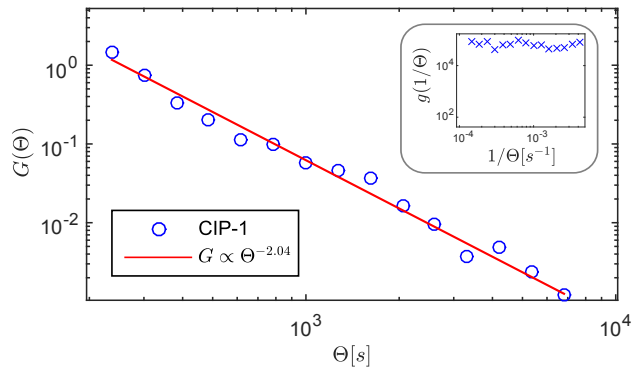}
	\caption{(color online) Burst time distribution $G(\Theta)$. The scaling (red line) corresponds to $G(\Theta) \propto \Theta^{-\gamma}$ with $\gamma=2.04 \pm 0.15$. In the inset we show the nearly uniform distribution $g(1/\Theta)$.}
	\label{fig:thetadist}
\end{figure}

Let us now focus on the distribution $G(\Theta)$ of time intervals $\Theta$ between two successive increments in the imposed pressure during which invasion bursts have occurred. Fig.~\ref{fig:thetadist} shows this distribution, produced for all bursts with $\Theta>120s$, a cutoff related to the minimum time difference for proceeding the image analysis used in the feedback mechanism \cite{moura2015}. It scales as $G(\Theta) \propto \Theta^{-\gamma}$ with $\gamma=2.04 \pm 0.15$ (the exponent was also computed via MLE \cite{clauset2009}). In the inset we show the distribution of inverse intervals $g(1/\Theta)$, which is nearly uniform, since it is related to $G(\Theta)$ by $g(1/\Theta)=G(\Theta) \Theta^2 \propto \Theta^{2-\gamma}$. The uniformity of $g(1/\Theta)$ will play an important role further on in the modeling of the pressure fluctuations PSD.

\section{Connection between the burst size and time distributions}
We consider now the link between the burst size distribution $N(n)$ shown in Fig.~\ref{fig:burstslocaldist} and the burst time distribution $G(\Theta)$ in Fig.~\ref{fig:thetadist}. Let $\dot{A}=s/\Theta$ denote the average growth rate of a burst of area $s$ during the time interval $\Theta$. This corresponds to an external perimeter growth \cite{feder1988,stauffer1994}, therefore $\dot{A} \propto u l_e$ where $u$ is a characteristic front speed (set by the Darcy law and the characteristic capillary pressure) and $l_e$ is the external perimeter, related to the linear size across a cluster $l$ as $l_e\propto l^{D_e}$. Since $s\propto l^{D}$, we have

\begin{equation}
s/\Theta = \dot{A} \propto u l_e \propto s^{D_e/D} \implies \Theta \propto s^\beta \:,
\label{eq:theta}
\end{equation}
with $\beta = 1-D_e/D$. The distributions of $s$ and $\Theta$ are linked by $\left|G(\Theta)d\Theta\right| = \left|p(s)ds\right|$ and since the area $s$ of a burst is proportional to its number of pores $n$ (see Fig.~\ref{fig:burstslocaldist}), it follows that $p(s)\propto s^{-\tau}$. Therefore,

\begin{equation}
G(\Theta)\propto s^{-\tau}ds/d\Theta \implies G(\Theta)\propto \Theta^{-\gamma} \:,
\label{eq:Gtheta}
\end{equation}
with $\gamma=\left(\tau-1+\beta\right)/\beta = \left(\tau-D_e/D\right)/\left(1-D_e/D\right)$. Using the literature values $\tau = 1.3$ \cite{martys1991,stauffer1994}, $D_e=1.33$ and $D=1.82$ \cite{stauffer1994}, we find $\gamma = 2.11$, quite close to the measured value $\gamma=2.04$ seen in Fig.~\ref{fig:thetadist}. As an immediate consequence of Eq.~(\ref{eq:Gtheta}), the distribution of inverse intervals scales as $g(1/\Theta) \propto \Theta^{-\eta}$, with $\eta=\gamma-2=\left(\tau-2+D_e/D\right)/\left(1-D_e/D\right)$. Using the literature values above we find $\eta=0.11$, which is in agreement with the experimentally observed value $\eta=\gamma-2=0.04 \pm 0.15$. These theoretical considerations explain the nearly uniform distribution observed in the inset of Fig.~\ref{fig:thetadist}.

\section{Fluctuations in the measured pressure signal}
\begin{figure}
	\centering
	\includegraphics[width=\linewidth]{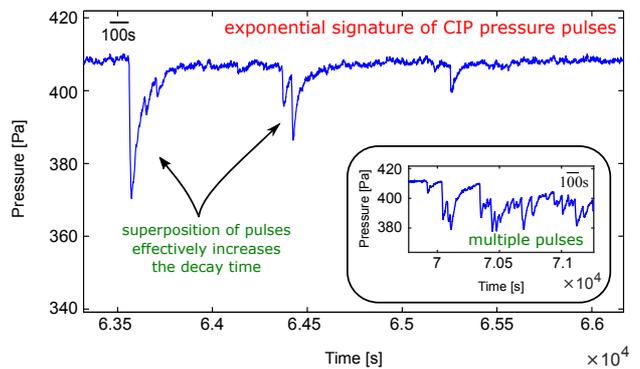}
	\caption{(color online) Typical exponential relaxation signature of pressure pulses. A pulse can trigger others and even give rise to a large avalanche (shown in the inset).}
	\label{fig:exponentialbusts}
\end{figure}
Next, we analyze the fluctuations in the pressure signal, following the pore invasion events. In Fig.~\ref{fig:exponentialbusts}, we show the typical pressure signature in a CIP experiment. The observed pressure pulses present a characteristic exponential relaxation. We also observe that a pulse can trigger others and even give rise to large avalanches with the invasion of several pores. A pulse can be divided into two phases: an initial fast drop in the capillary pressure $p_c$ and a slower exponential relaxation back to the pressure level $\rho g h$ set externally (see Fig.~\ref{fig:diagram}). The fast drop in $p_c$ occurs as the liquid is displaced (following the invasion of one or more pores) and subsequently redistributed to the surrounding menisci, causing a back-contraction of the interface \cite{maloy1992,furuberg1996,haines1930}. The relaxation phase occurs as the liquid-air interface readjusts itself inside the available pore-throats and the liquid volume displaced from the pores flows out of the model. The fluid motion sets in viscous pressure drops which are reflected in the measured pressure, as seen in Fig.~\ref{fig:exponentialbusts}. These drops occur (see Fig.~\ref{fig:diagram}): 1) in the porous medium itself, 2) in the filter at the model's outlet and 3) in the external tubing (the numbers are in correspondence with Fig.~\ref{fig:diagram}). The height difference $h$ between the surface of the liquid in the reservoir and the model level accounts for a hydrostatic component $\rho g h$. Adding these contributions and assuming that the flow is governed by Darcy's equation, we have
\begin{equation}
p_w-u\frac{\mu L_1}{k_1}-u\frac{\mu L_2}{k_2}-u\frac{S_1 \mu L_3}{S_3 k_3} + \rho g h = p_0 \:,
\label{eq:S1}
\end{equation}
where $p_w$ is the pressure in the wetting phase (liquid) just after the liquid-air interface, $u$ is the average Darcy velocity of the flow in the porous network, $\mu=\rho \nu$ is the liquid's dynamic viscosity, $L_i$ and $k_i$ with $i=\left\{1,2,3\right\}$ are the length and permeability respectively of the porous network, filter and the tubing and $S_1$ and $S_3$ are the respective cross sections of the model and the tubing. Since the capillary pressure across the liquid-air interface is $p_c=p_{nw}-p_w=p_0-p_w$, Eq.~(\ref{eq:S1}) becomes

\begin{equation}
p_c + uR - \rho g h = 0 \:,
\label{eq:S2}
\end{equation}
where

\begin{equation}
R = R_1 + R_2 + R_3 \implies R = \frac{\mu L_1}{k_1} + \frac{\mu L_2}{k_2} + \frac{S_1 \mu L_3}{S_3 k_3} \:,
\label{eq:resistence}
\end{equation}
is equivalent to an effective resistance to the flow. The volumetric flux in a pore is $dV/dt = u a^2/\phi$, where $a$ is a characteristic pore length scale (for example the bead diameter) and  $\phi$ is the porosity of the model. By introducing the concept of a capacitive volume $\kappa=dV/dp_c$ (used first in Ref.~\cite{maloy1992}), where $dV$ is the liquid volume displaced from a pore throat in response to a change $dp_c$ in capillary pressure, we have

\begin{equation}
\frac{dV}{dt} = \frac{u a^2}{\phi} \implies u = \frac{\kappa \phi}{a^2} \frac{dp_c}{dt} \:.
\label{eq:S3}
\end{equation}
Plugging this equation into Eq.~(\ref{eq:S2}),

\begin{equation}
\frac{\kappa \phi R}{a^2} \frac{dp_c}{dt} + p_c - \rho g h = 0 \implies p_c(t)= \rho g h + C e^{-t/t_c} \:,
\label{eq:S5_2}
\end{equation}
thus producing the exponential behavior seen in Fig.~\ref{fig:exponentialbusts}. $C=p_c(0)-\rho g h<0$ is a constant associated to how much the capillary pressure decreases during the invasion of a set of pores before it starts to rise again. The characteristic time scale of the exponential decay is
\begin{equation}
t_c= \frac{\kappa \phi R}{a^2} \:.
\label{eq:tc}
\end{equation}

The invasion of one pore quite frequently triggers the invasion of others, in such a manner that before an exponential pulse decays completely, another one is seen in the pressure signal, see Fig.~\ref{fig:exponentialbusts}. This mechanism delays the complete relaxation of the pressure, effectively increasing the decay time from $t_c$ to $t^*\geq t_c$. Indeed, if this relaxation-delaying mechanism was absent, the burst time distribution $G(\Theta)$ shown in Fig.~\ref{fig:thetadist} should be peaked around the value $\Theta = t_c$. Since we have shown that $G(\Theta) \propto \Theta^{-\gamma}$ we expect the effective exponential decay time $t^*$ to follow the same distribution and, in particular, the effective decay rate $\lambda=1/t^*$ should be uniformly distributed in an interval $[\lambda_{min}, \lambda_{max}]$ following the same distribution as $1/\Theta$ (see inset of Fig.~\ref{fig:thetadist}). $\lambda_{max}$ is related to the minimum decay time $t^*$, i.e., $\lambda_{max}=1/t_c$ and we will consider $\lambda_{min}=0$ for convenience. Later on we will show that the distribution of decay rates has a crucial impact on the power spectrum of the pressure signal.

\section{Pressure signal PSD}
\begin{figure}
	\centering
	\includegraphics[width=\linewidth]{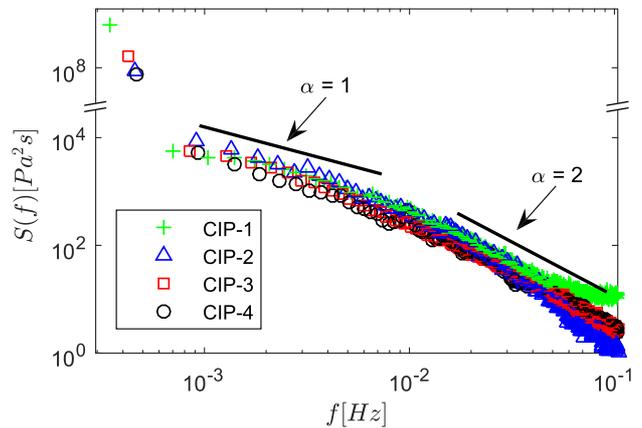}
	\caption{(color online) Power spectral density comparison for CIP experiments (model's numbers in the legend). Guide-to-eye lines are shown for the scaling $S(f)\propto~f^{-\alpha}$, with $\alpha = 1$ for lower frequencies and $\alpha = 2$ for intermediate frequencies.}
	\label{fig:psd4cp}
\end{figure}

In Fig.~\ref{fig:psd4cp} we show the power spectral density (PSD) associated to the pressure signal for the CIP experiments. The PSD $S=S(f)$ was computed for all experiments using the Welch method \cite{welch1967}. We have noticed the existence of a $1/f$ scaling regime (flicker/pink noise) for lower frequencies, followed by a crossover and a $1/f^2$ scaling regime (brown noise) for intermediate frequencies. For higher frequencies, another crossover follows and a region independent of $f$ is seen (white noise associated with fluctuations in the pressure sensor and unimportant to our analysis). We see from Fig.~\ref{fig:psd4cp} that the scaling properties of the power spectrum, in particular the occurrence of $1/f$ noise, seem to remain unchanged despite the changes in both sample dimensions and pore-size distribution (the samples were rebuilt before each experiment, thus changing the pore-size distribution \cite{moura2015}). The $1/f$ regime is associated with events having frequency $f<10^{-2}Hz$, or alternatively, periods $T>100s$. From Fig.~\ref{fig:exponentialbusts}, we see that this corresponds to the characteristic time intervals between the pressure pulses, thus indicating that they are associated with the presence of the $1/f$ scaling in the PSD.

\section{Analytical modeling of the pressure signal and PSD scaling explanation}
The non-trivial scaling of the CIP power spectral density can be explained by the following mathematical framework, which is an adaptation of an argument proposed in \cite{ziel1950} to explain a similar $1/f$ to $1/f^2$ transition in the very first reported observation of $1/f$ noise \cite{johnson1925} (see also \cite{bernamont1937} and \cite{milotti2002}). Apart from a nearly constant offset, the pressure signal can be modeled as a train of exponentially decaying pulses located at randomly distributed discrete times $t_j$,

\begin{equation}
p_\lambda(t)=\sum_jAH(t-t_j) e^{-\lambda(t-t_j)} \:,
\label{eq:train}
\end{equation}
where $\lambda>0$ and $A<0$ are initially taken to be constants (the characteristic decay rate and amplitude of the pulses) and $H(t-t_j)$ is the Heaviside step function, i.e., $H(t-t_j)=0$ if $t<t_j$ and $H(t-t_j)=1$ if $t\geq t_j$. Let $P_\lambda(f)$ be the Fourier transform of $p_\lambda(t)$. The PSD $S_\lambda(f)$ is

\begin{equation}
S_\lambda(f)=\lim_{T \to \infty}\frac{1}{T} \left\langle \left|P_\lambda(f)\right|^2\right\rangle = \frac{A^2 r}{\lambda^2+4 \pi^2 f^2} \:,
\label{eq:lorentzian}
\end{equation}
where $r$ is the average rate of occurrence of pulses and the brackets are the expected value operator (since in practice one does not have access to an ensemble of measurements, we have employed Welch's method \cite{welch1967} to estimate the PSD, which is based on the concept of a periodogram \cite{press1992}). The PSD shown in Eq.~(\ref{eq:lorentzian}) is a Lorentzian curve which is approximately constant for lower frequencies ($f \ll \lambda / 2 \pi$) and decays as $1/f^2$ for higher frequencies ($f \gg \lambda / 2 \pi$).

\begin{figure}
	\centering
	\includegraphics[width=\linewidth]{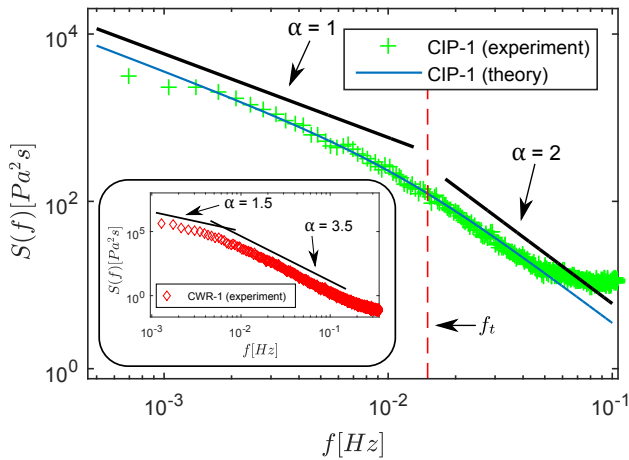}
	\caption{(color online) Comparison between theoretical prediction and experiments. The analytical result (thin blue line) is given by Eq.~(\ref{eq:lorentzian2simplified2}), where $f_t=1.5\cdot10^{-2}Hz$ (vertical dashed red line) and $A^2 r = 1.5\:Pa^2/s$. The analytical prediction match the experimental measurements (green crosses, experiment CIP-1) well. On the inset we show the PSD for experiment CWR-1.}
	\label{fig:theorycompare}
\end{figure}

A model with a single constant decay rate $\lambda$ cannot incorporate the $1/f$ region but, as previously argued, we expect $\lambda$ to follow the uniform distribution $\xi(\lambda)=1/\lambda_{max}$ in the interval $[0, \lambda_{max}]$. Taking this distribution into account and writing $\lambda_{max}=2\:\pi\:f_t$, we have

\begin{equation}
S(f)=\int_{0}^{\lambda_{max}}\!\!\! S_\lambda(f) \xi(\lambda) d\lambda = \frac{A^2 r}{4 \pi^2 f_t f} \arctan\left(\frac{f_t}{f}\right) \:.
\label{eq:lorentzian2simplified2}
\end{equation}
Eq.~(\ref{eq:lorentzian2simplified2}) has the asymptotic behavior
\begin{equation}
S(f) = \begin{cases}
\frac{A^2 r}{8 \pi f_t}\frac{1}{f} &\:\:\text{if $f \ll f_t$}\\
\frac{A^2 r}{4 \pi^2}\frac{1}{f^2} &\:\:\text{if $f \gg f_t$}
\label{eq:lorentzian2simplified3}
\end{cases} \:,
\end{equation}
thus presenting the $1/f$ to $1/f^2$ transition observed in the experiments. The transition frequency $f_t$ in experiment CIP-1 is roughly $f_t=1.5\cdot10^{-2}Hz$ (see Fig.~\ref{fig:psd4cp}). By using the constant $A^2 r$ as a fitting parameter we can compare the measured PSD with the theoretical prediction in Eq.~(\ref{eq:lorentzian2simplified2}). Fig.~\ref{fig:theorycompare} shows the resulting comparison produced using $A^2 r = 1.5\;\!Pa^2/s$. The dashed red vertical line marks the transition frequency $f_t$. The analytical result reproduces the experimental findings very well, scaling as $1/f$ for $f \ll f_t$ and as $1/f^2$ for $f \gg f_t$. Indeed, this theory not only captures the $1/f$ and $1/f^2$ domains but also fits the data well for the crossover region between these domains.

The transition frequency $f_t$ can be estimated using Eq.~(\ref{eq:tc}) and the resistance $R$ from Eq.~(\ref{eq:resistence}). As a first order approximation, let us consider only the contribution to $R$ from the term $R_1$ relative to the resistance in the porous medium itself. Using $\mu=\rho \nu = 5.1 \cdot 10^{-2} Pa.s$, $L_1 = 0.27 m$, $a=10^{-3}m$, $\kappa = 1.1 \cdot 10^{-12} m^3/Pa$ (from Ref.~\cite{maloy1992}), $k_1=1.6 \cdot 10^{-9}m^2$ and $\phi=0.63$ (both measured in a similar model in Ref.~\cite{grunde2004}), we find

\begin{equation}
f_t = \frac{1}{2 \pi t_c} = \frac{k_1 a^2}{2 \pi \kappa \phi \mu L_1}\implies f_t \approx 2.6\cdot10^{-2} Hz \:,
\label{eq:S7}
\end{equation}
not far from the transition frequency $f_t =1.5\cdot10^{-2}Hz$ shown in Fig.~\ref{fig:theorycompare}. The overestimation comes from the terms $R_2$ and $R_3$ in Eq.~(\ref{eq:resistence}), ignored in the calculation above.

Finally, notice also the existence of a single isolated point in the very low frequency part of the PSD, falling far from the scaling region (extreme left for all experiments in Fig.~\ref{fig:psd4cp}). This point is not an outlier in the data: its existence signals the very slow positive drift of the pressure signal, which occurs since the capillary pressure has to increase to allow the invasion of narrower pores \cite{maloy1992,moura2015}.

\section{Comparison with a system driven under a CWR boundary condition}
In order to test the effect of the boundary conditions in the PSD, we have run a controlled withdrawal rate (CWR) experiment using model (1). The resulting PSD is shown in the inset of Fig.~\ref{fig:theorycompare}. The PSD still presents an interesting scaling, but with different scaling regimes: $1/f^{1.5}$, for lower frequencies, and $1/f^{3.5}$, for intermediate frequencies. The $1/f$ region is only observed for systems driven under the CIP boundary condition. The fact that the exponents for CWR differ from CIP is not surprising, since the pressure relaxation in that case is no longer exponential, but linear, see Ref.~\cite{maloy1992}.

\section{Connection between the measured pressure and the capillary pressure}
The pressure sensor measures the difference between the pressure in the air and the liquid at the outlet, i.e., $p_m = p_{nw}-p_w^{out}$. The measured signal is not exactly the capillary pressure $p_c=p_{nw}-p_w$ across the liquid-air interface, since $p_w \neq p_w^{out}$ given that viscous losses occur between the liquid-air interface and the outlet, thus generally making $p_w > p_w^{out}$. Those losses occur in the porous medium itself and in the filter at the outlet of the model (numbers (1) and (2) in Fig.~\ref{fig:diagram}). The connection between $p_m$ and $p_c$ is $p_m = p_c +u\left(R_1+R_2\right)$, where $R_1$ and $R_2$ are the resistance terms from the porous network and the filter. Using Eqs.~(\ref{eq:S3}) and (\ref{eq:S5_2}), we have

\begin{equation}
p_m = \rho g h + C\left(1-\frac{R_1+R_2}{R_1+R_2+R_3}\right) e^{-t/t_c}\:.
\label{eq:S9}
\end{equation}
Therefore, by comparing Eqs.~(\ref{eq:S5_2}) and (\ref{eq:S9}), we see that $p_m$ differs from $p_c$ only in the amplitude of the pulses, but not in their characteristic exponential decay. Since our analysis depended only on the distribution of the decay rates, the differences between $p_m$ and $p_c$ are not crucial.

\section{Further generalizations of the PSD analytical framework}
One possible generalization of the model would be to consider a system with a distribution of amplitudes $A$ instead of a single value (as we might expect from Fig.~\ref{fig:exponentialbusts}). In this case the scaling properties of the PSD would still be left unchanged but the constant $A^2$ in Eq.~(\ref{eq:lorentzian2simplified2}) and (\ref{eq:lorentzian2simplified3}) would be replaced by the expected value of $A^2$. Another possibility would be to consider a distribution for $\lambda$ of the form $\xi(\lambda)\propto\lambda^{-\delta}$. Here the $1/f^2$ region is still left unchanged but the $1/f$ scaling is changed to $1/f^{(1+\delta)}$ \cite{butz1972}. As previously noted, the distribution of decaying rates $\lambda$ is the crucial figure behind the $1/f$ scaling.

\section{Conclusions}
We have analyzed the burst dynamics from slow drainage experiments in porous media. We showed that this dynamics presents many features commonly associated to critical systems. Intermittent bursts of activity were observed over many time and length scales and a theoretical expression for their size distribution scaling, Eq.~(\ref{eq:tau}), was verified experimentally. The pressure signal of the invasion presented an interesting PSD scaling, with a $1/f$ scaling region which further transitions to $1/f^2$ in the case of the CIP boundary condition. We have employed an analytical framework \cite{ziel1950} which satisfactorily reproduces the scaling properties of the PSD. The derivation of closed expressions relating the pressure signal PSD to properties of the porous medium and the fluids can lead to new techniques for indirectly probing such systems. For example, if one has access to the PSD only and not to the full pressure signal, the transition frequency $f_t$ can still be measured and information on the ratio $k_1/\phi$ between the permeability and the porosity of the medium can be found via Eq.~(\ref{eq:S7}). If the PSD and $f_t$ are known, Eq.~(\ref{eq:lorentzian2simplified2}) can be fitted to measure the product $A^2 r$ between the amplitudes and rate of occurrence of bursts. 

\section*{\large{Acknowledgments}}
We acknowledge the support from the University of Oslo, University of Strasbourg, the Research Council of Norway through its Centre of Excellence funding scheme with project number 262644, the CNRS-INSU ALEAS program and the EU Marie Curie ITN FLOWTRANS network.

\bibliographystyle{ieeetr}
\bibliography{bib}

\end{document}